\documentclass{article}[11pt]
\usepackage[utf8]{inputenc}
\usepackage{amsmath,amssymb,bbm}

\newcommand{\argmin}{\mathop{\arg\min}}
\newcommand{\bs}{\boldsymbol}
\newcommand{\diff}{\mathrm{d}}

\newcommand{\interval}{h}

\usepackage{algorithm}
\usepackage{algorithmicx}
\usepackage{algpseudocode}
\usepackage{bbm}

\usepackage{subcaption}

\usepackage[whole]{bxcjkjatype}

\usepackage{graphicx}[dvipdfmx]

\usepackage{mdframed}
\usepackage{hyperref}
\hypersetup{
    colorlinks = true,
	citecolor=blue,
	linkcolor=blue
}

\usepackage{booktabs}

\usepackage{geometry}
\usepackage{authblk}

\usepackage{enumerate}
\usepackage{amsthm}
\theoremstyle{definition}

\newtheorem*{theorem*}{Theorem}

\newtheorem*{note*}{Note}
\newtheorem{ex}{Example}

\newtheorem{prop}{Proposition}

\usepackage{appendix}
\usepackage{fourier}

\newcommand{\setX}{\mathcal{X}}

\newcommand{\interaction}{\mathcal{I}}

\title{Autoregressive with Slack Time Series Model for Forecasting a Partially-Observed Dynamical Time Series}
\author[1,2]{Akifumi Okuno\thanks{okuno@ism.ac.jp}}
\author[3]{Yuya Morishita\thanks{morishita.yuya.7x@kyoto-u.ac.jp}}
\author[4]{Yoh-ichi Mototake\thanks{y.mototake@r.hit-u.ac.jp}}
\affil[1]{The Institute of Statistical Mathematics}
\affil[2]{RIKEN Center for Advanced Intelligence Project}
\affil[3]{Kyoto University}
\affil[4]{Hitotsubashi University}

\date{\empty}

\begin{document}

\maketitle

\begin{abstract}
This study delves into the domain of dynamical systems, specifically the forecasting of dynamical time series defined through an evolution function. Traditional approaches in this area predict the future behavior of dynamical systems by inferring the evolution function. However, these methods may confront obstacles due to the presence of missing variables, which are usually attributed to challenges in measurement and a partial understanding of the system of interest. To overcome this obstacle, we introduce the autoregressive with slack time series (ARS) model, that simultaneously estimates the evolution function and imputes missing variables as a slack time series. Assuming time-invariance and linearity in the (underlying) entire dynamical time series, our experiments demonstrate the ARS model's capability to forecast future time series. From a theoretical perspective, we prove that a 2-dimensional time-invariant and linear system can be reconstructed by utilizing observations from a single, partially observed dimension of the system.
\end{abstract}


\section{Introduction}
\label{sec:introduction}
Notwithstanding its difficulty, forecasting of the evolution of intricate non-linear dynamical time series has been in the spotlight in various scientific fields~\cite{strogatz2001nonlinear,jackson2015applications}. 
A plausible approach to forecasting the evolution is to isolate the non-linear estimation problem into (i) learning non-linear representations by applying highly non-linear functions such as deep neural networks~\cite{Goodfellow-et-al-2016}, and (ii) estimating its evolution with simple linear models. 
One example is a reservoir computing (RC)~\cite{jaeger2001echo,jaeger2002adaptive}. 
RC first randomly specifies a state in the reservoir layer in recurrent neural network~\cite{rumelhart1986}, and optimizes the weights only in the output layer; RC corresponds to non-linearly transform its input (in the reservoir layer) and trains a simple linear prediction model (in the output layer). 
It has been reported that such a simple combination of the non-linear representation learning and the linear estimation is effective to forecasting the evolution of intricate dynamical series~\cite{tanaka2019recent}. 
Effectiveness of the simple combination is not limited to RC; applying a linear model to the non-linear representation in more general deep neural network is also regarded as a solid forecasting strategy~\cite{lusch2018deep}.

Unfortunately, however, partial degrees of freedom corresponding to several state variables are not observed in some practical situations~\cite{lucor2022simple,cheng2023machine}. 
There could be a variety of reasons for missing observations: 
it would be caused by the difficulty of measurement, 
it would be caused by the immature understanding of the system of interest, and so forth. 
Generally speaking, it is quite difficult to find and identify all the related state variables in the system in real world situations. 
To address the issue, studies in line with dynamical systems have considered and substituted the delay embedding~\cite{packard1980geometry, alligood1998chaos}, with the aid of delay embedding theorem (also known as Takens' theorem)~\cite{takens1981detecting}. 
While the delay embedding theorem indicates that the (topological) attractor structure of the entire dynamical system is recovered by the delay embedding of the partial observations, i.e., this theorem provides a ``rough estimate'' of the entire system, it does not intend to provide any accurate forecasting. 
Therefore, for more accurate forecasting purposes, \cite{gottwald2021combining} leverages the random feature maps of the delay embedding, \cite{ouala2020learning} applies a deep neural network to the delay embedding, and \cite{ouala2023bounded} proposes estimating the missing variables themselves directly by a simple transformation of the observed variables. 
Therein, neural ordinal differential equation (neural ODE)~\cite{chen2018neural} is used to model the evolution of the time series. 
In line with these approaches, forecasting of the evolution of a partially-observed dynamical time series with the estimation of the missing variables has been actively studied recently~\cite{cheng2023machine}.

\begin{figure}[!ht]
\centering
    \includegraphics[width=0.5\textwidth]{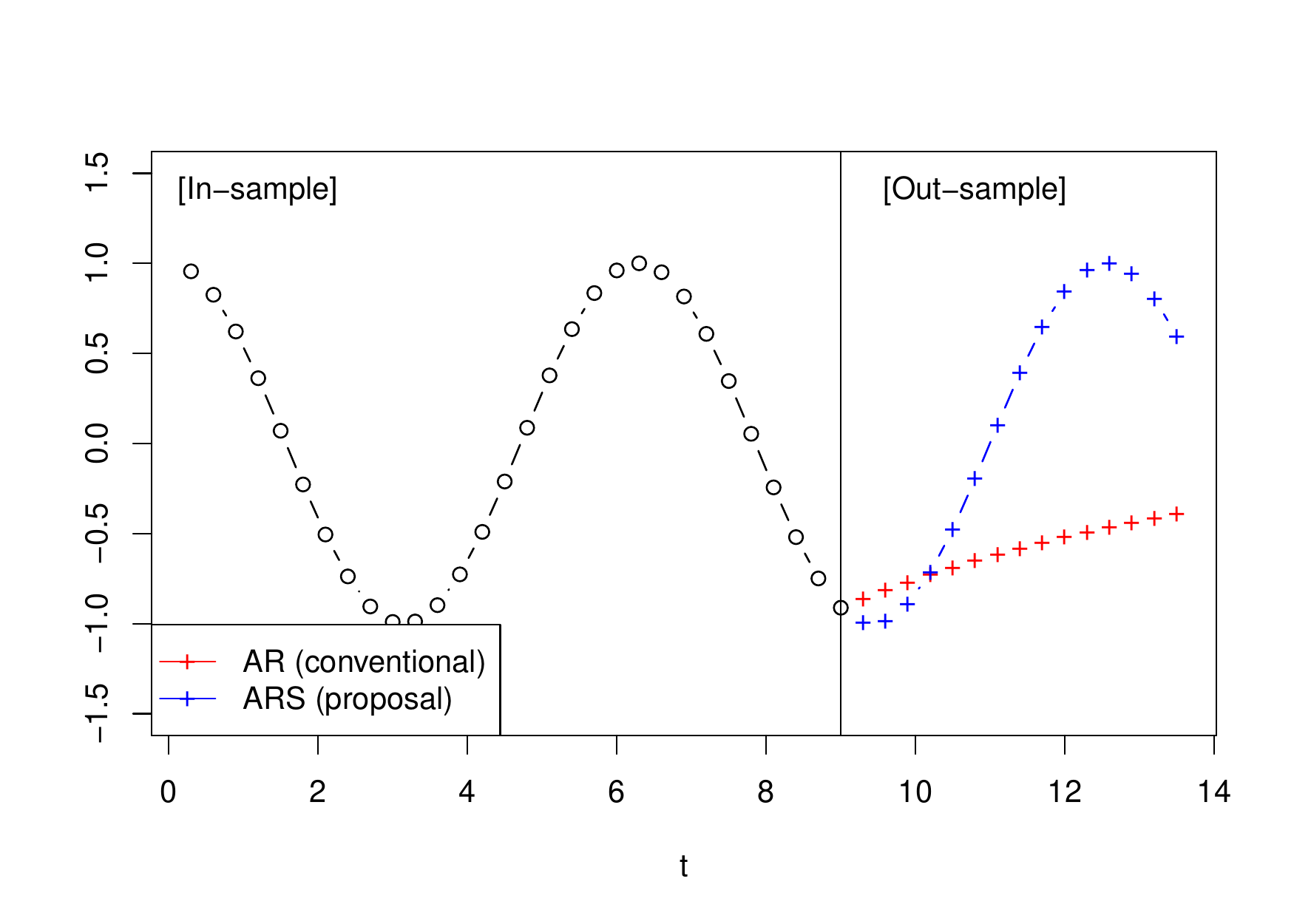}
    \vspace{-1em}
    \caption{Time series forecasting with (red) conventional AR model shown in Equation~\eqref{eq:AR_observed} and (blue) proposed ARS model shown in Equation~\eqref{eq:ARS}. See Section~\ref{subsec:ARS} for further details of the experiment.}
    \label{fig:illustration}
\end{figure}

Recent research has leveraged deep neural networks to reconstruct the full dynamics of systems, but the complexity of training these networks highlights the appeal of simpler, more manageable models as alternatives. Consider a straightforward scenario where a $2$-dimensional time series exhibits circular motion, yet only the first dimension (corresponding to a cosine curve) is observable. In such cases, while the complete dynamics adhere to a simple, time-invariant linear system at regular discrete time points (that are considered in many literature; see, e.g., \cite{satheesh2023antiwindup} and \cite{satheesh2024design}), the observed cosine function exhibits time-variance. This discrepancy suggests that partial observations may appear to follow ostensibly complex patterns, even when the underlying dynamics are fundamentally simpler. 
Motivated by this fact, we introduce the autoregressive with slack time series (ARS) model, designed to estimate the evolution function and impute missing variables using a slack time series simultaneously. By assuming the entire underlying dynamical time series is time-invariant and linear, our experiments validate the efficacy of the ARS model. Theoretically, we demonstrate that it is possible to reconstruct a 2-dimensional, time-invariant, and linear system using observations from just one partially observed dimension, as depicted in Figure~\ref{fig:illustration}.

Organization of this paper is as follows: 
Section~\ref{subsec:symbols} describes the symbols and notations used throughout this paper, 
Section~\ref{sec:preliminaries} describes the preliminaries, 
Section~\ref{sec:ARS} describes the proposed ARS model, 
Section~\ref{sec:experiments} describes the numerical experiments, and 
Section~\ref{sec:discussions_conclusions} describes the remaining discussions and conclusions. 
Particularly, the discussion section includes a theoretical guarantee for the ARS model prediction with a simple setting, and relations to classical state-space models and higher-order autoregressive~(AR) models.

\subsection{Symbols and notations}
\label{subsec:symbols}
This section provides a summary of the symbols used throughout this paper. The symbol $t \ge 0$ represents time, and $x=x(t)$ signifies a state at time $t$, which takes on a value within the non-empty state space $\mathcal{X} \subset \mathbb{R}^d$. 
As long as $x(t)$ follows a dynamical system, the temporal state $x(t)$ is also referred to as the dynamical time series. 
Time $t$ is discretized into small intervals $h>0$, such that $t=jh$ for $j=1,\ldots,n$. At each discretized time point $t=jh$, the state $x(jh) \in \mathbb{R}^d$ is split into an observed component $z(jh) \in \mathbb{R}^r$ and a missing (hidden) component $z^{\dagger}(jh) \in \mathbb{R}^s$, resulting in $x(jh)=(z(jh),z^{\dagger}(jh))$. The sequence $\{z_j^{\dagger}\}_{j=1}^{n}$ represents a slack time series, where each element $z_j^{\dagger}$ acts as a substitute for the missing component $z^{\dagger}(jh)$. This study aims to estimate the slack time series as $\hat{z}_j^{\dagger}$, with the completed time series expressed as $\hat{x}^{\ddagger}(jh)=(z(jh),\hat{z}_j^{\dagger})$ for $j=1,2,\ldots,n$.

\section[Preliminaries]{Preliminaries}
\label{sec:preliminaries}

This section provides preliminaries. 
More specifically, 
Section~\ref{subsec:dynamical_system} describes the dynamical system, 
Section~\ref{subsec:autoregressive_model} describes the autoregressive model, and 
Section~\ref{subsec:problem_setting} describes the problem setting considered in this study.

\subsection{Dynamical system}
\label{subsec:dynamical_system}

Let $d \in \mathbb{N}$. 
A \emph{dynamical system} is a pair $(\setX,\varphi)$, where $\setX \subset \mathbb{R}^d$ is a non-empty set called \emph{state space} and $\varphi:\mathbb{R}_{\ge 0} \times \setX \to \setX$ denotes an \emph{evolution function} satisfying
\begin{align*}
    \varphi(t', \varphi(t, \bs x)) = \varphi(t+t', \bs x), 
    \quad 
    \varphi(0,\bs x)=\bs x 
\end{align*}
for all $t,t' \ge 0$ and $\bs x \in \setX$. 
Intuitively speaking, $\varphi(t,\bs x) \in \setX$ represents the state evolved from $\bs x \in \setX$ during the period of time $t \ge 0$. 
As $\bs x=\bs x(t) \in \setX$ depends on the time $t \ge 0$, we call $\bs x(t)$ as a \emph{(dynamical) time series} herein. 
Typically, the dynamical system can be specified by a differential equation 
\[
\frac{\diff \bs x(t)}{\diff t}
=
\frac{\diff \varphi(t',\bs x(t))}{\diff t'} \bigg|_{t'=0}
=
f_{t}(\bs x(t))
\]
equipped with the \emph{time derivative} $f_t:\setX \to \mathcal{T}$ defined with some set $\mathcal{T} \subset \mathbb{R}^{d}$. 
The time derivative $f_t$ is called \emph{time-invariant} if $f_t$ is independent of the time $t$ (herein, $f$ represents such a time-invariant derivative); 
an example of such a time-invariant dynamical system (with $d=3,\bs x=(x_1,x_2,x_3)$) is the Lorenz system~\cite{lorenz1963deterministic} $f(\bs x)=(-\alpha x_1+\alpha x_2, -x_1 x_3+ \beta x_1 - x_2, x_1 x_2 - \gamma x_3)$ for some constants $\alpha,\beta,\gamma \in \mathbb{R}$. See, e.g., \cite{strogatz2001nonlinear} for the long-standing history of the dynamical systems.

While the above dynamical system considers the evolution of the series in continuous time, in practice, the evolving states are observed only for the discrete-time $t=\interval, 2\interval, \ldots, n\interval$; with sufficiently short period of time $\interval>0$, the evolution of the time-invariant dynamical system can be approximated by a first-order Taylor expansion 
\begin{align}
    \bs x(t+\interval)
    &=
    \varphi(\interval, \bs x(t)) \nonumber\\
    &=
    \varphi(0,\bs x(t)) + \interval \frac{\diff \varphi(t',\bs x(t))}{\diff t'} \bigg|_{t'=0} + O(\interval^2) \nonumber \\
    &=
    \bs x(t) + \interval f(\bs x(t)) + O(\interval^2).
    \label{eq:short_period_approximation}
\end{align}
$O(h^2)$ denotes the term smaller than $h^2$ in the limit $h \searrow 0$. 
By assuming the linearity in the time derivative $f$, i.e., $f(\bs x)=A\bs x$ for some matrix $A \in \mathbb{R}^{d \times d}$, the approximation shown in Equation~\eqref{eq:short_period_approximation} indicates that the state evolved for the short period of time $\interval \ge 0$ is approximated by a simple linear transformation
\begin{align}
    \bs x(t+\interval)
    =
    B\bs x(t) + O(\interval^2)
\label{eq:AR_dynamical_system}
\end{align}
for some matrix $B=B(\interval):=I+\interval A \in \mathbb{R}^{d \times d}$ defined with the $d \times d$ identity matrix $I$. 
The short period approximation of the time-invariant dynamical system shown in Equation~\eqref{eq:AR_dynamical_system} leads to the autoregressive model described in the following Section~\ref{subsec:autoregressive_model}.

While the discussions below consider only the linear time derivative $f$ for simplicity, 
dynamical systems equipped with even non-linear (and continuous) time derivative $f$ can be approximated by a polynomial extension. Also see the discussion in Appendix~\ref{app:interaction}.

\subsection{Autoregressive model}
\label{subsec:autoregressive_model}

With a positive integer $n$, assume that the evolving states $\bs x(t)$ are observed at the discrete timepoints $t=\interval, 2\interval, 3\interval,\ldots, n\interval$. 
For modeling the short period approximation of the dynamical system shown in Equation~\eqref{eq:AR_dynamical_system}, we may employ an \emph{autoregressive~(AR)} model of order $1:$
\begin{align}
    \hat{\bs x}((j+1)\interval) = \hat{B} \bs x(j\interval) \quad (j=1,2,\ldots,n-1),
    \label{eq:vanilla_AR}
\end{align}
where $\hat{B} \in \mathbb{R}^{d \times d}$ is a matrix typically estimated by minimizing the loss function 
\[
    \hat{B} := \argmin_{B \in \mathbb{R}^{d \times d}}
    \sum_{j=1}^{n-1}
    \|
        \bs x((j+1)\interval) - B \bs x(j\interval)
    \|_2^2.
\]
While this study considers the AR model of order $p=1$ (typically denoted by AR(1)) for simplicity, the AR(1) model can be straightforwardly extended to the AR model of higher order $p \in \mathbb{N}$: $\hat{\bs x}((j+1)\interval)=\sum_{k=1}^{p} \hat{B}_{k} \bs x((j+1-k)\interval)$, with the estimated matrices $\{\hat{B}_k\}_{k=1}^{p} \subset \mathbb{R}^{d \times d}$. See, e.g., \cite{mills1990time} and \cite{hamilton1994time} for details of the extensions, and Section~\ref{subsec:discussion_ARp} for further discussions.

\subsection{Problem setting}
\label{subsec:problem_setting}

Herein, assume that the evolving time series $\{\bs x(j \interval)\}_{j=1,2,\ldots,n}$ follows the time-invariant dynamical system (shown in Equation~\eqref{eq:AR_dynamical_system}) equipped with the linear time derivative $f$. 
The AR model shown in Equation~\eqref{eq:vanilla_AR} is expected to approximate the dynamics well. 
However, in some practical situations, not all the variables in the state $\bs x \in \mathcal{X}$ can be observed; we assume that the time series is divided into two time series of observed and missing (hidden) variables:
\[
    \bs x(j\interval) = (\bs z(j\interval),\bs z^{\dagger}(j\interval)) \quad (j=1,2,\ldots,n),
\]
where $\bs z(t) \in \mathbb{R}^{r}$ represents the time series of the observed variables (of interest) and $\bs z^{\dagger}(t) \in \mathbb{R}^{s}$ represents that of the missing variables, with $r,s \in \mathbb{N}$ satisfying $r+s=d$. 
As the time series $\bs z^{\dagger}(t)$ is missing, the conventional AR model can consider only the observed part in this setting, i.e., 
\begin{align}
    \hat{\bs z}((j+1)\interval) = \hat{C} \bs z(j\interval) \quad (j=1,2,\ldots,n-1),
    \label{eq:AR_observed}
\end{align}
where $\hat{C} \in \mathbb{R}^{r \times r}$ is a matrix typically estimated by minimizing the loss function 
$\sum_{j=1}^{n} \|\bs z((j+1)\interval) - C\bs z(j\interval)\|_2^2$.

\begin{figure*}[!ht]
\centering
\includegraphics[width=0.7\textwidth]{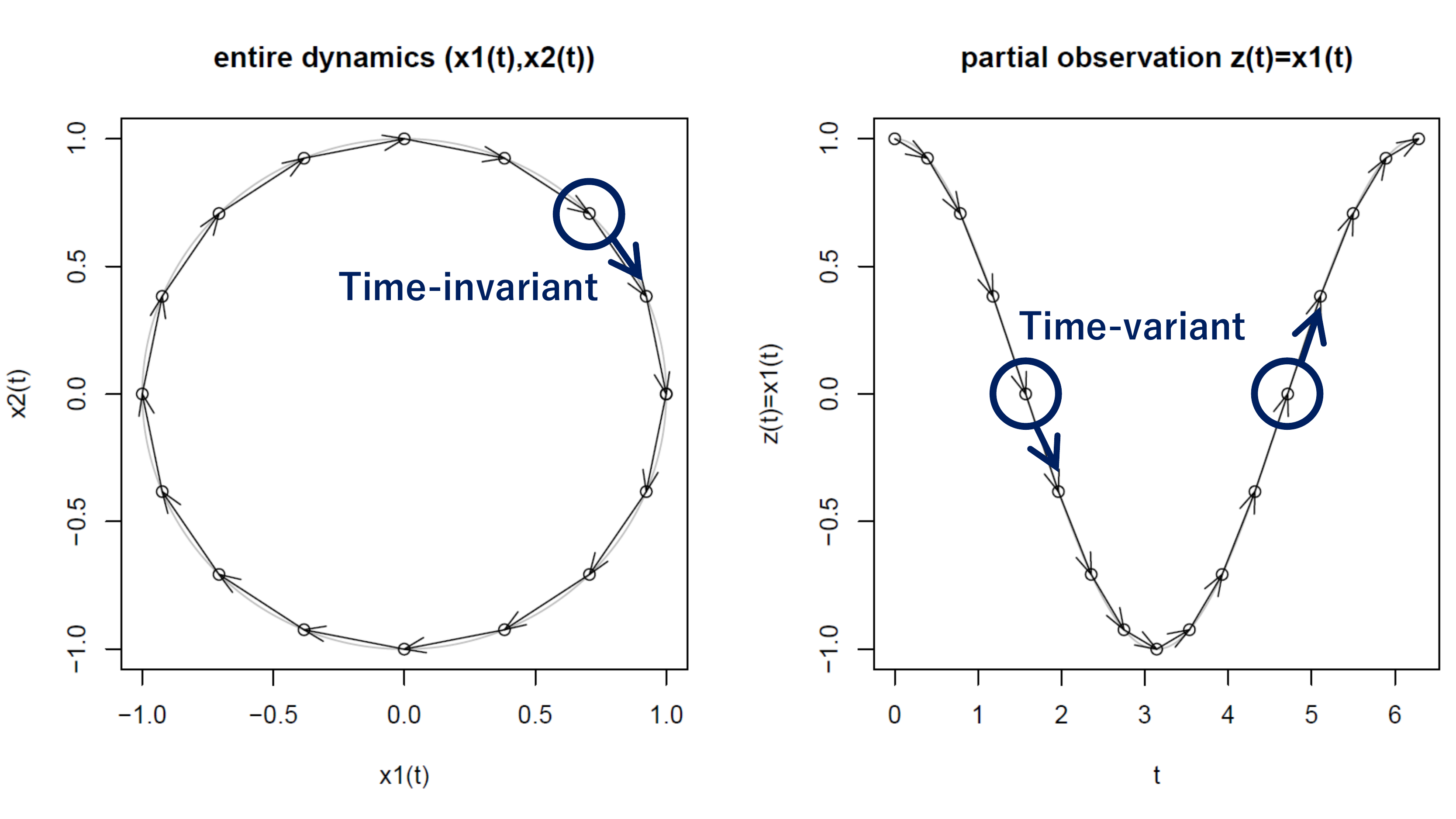}
\caption{Circular motion. While (left) the entire dynamics $\bs x(t)=(\cos t,\sin t)$ is time-invariant (as $\bs x(t+h)$ can be identified as a function of $\bs x(t)$), 
(right) the partial observation $z(t)=x_1(t)=\cos t$ is time-variant (as the next state of $x(t)=0$ can be both $x(t+h)>0$ and $x(t+h)<0$ depending on the current time $t$). 
}
\label{fig:circular_motion}
\end{figure*}

However, while the time-invariance of the system is considered important as also discussed in Section~\ref{subsec:dynamical_system}, evolution of the partial observation $\bs z(t)$ may lose the time-invariance property, even if the entire dynamics $\bs x(t)$ is time-invariant. 
A simple example is the circular motion; see Example~\ref{ex:circular_motion} and Figure~\ref{fig:circular_motion}. 

\begin{ex}
\label{ex:circular_motion}
Consider the case that $\bs x(t)$ follows a circular motion in $\mathbb{R}^2$, i.e., $\bs x(t)=(\cos t,\sin t)$, and assume that $r=s=1$, i.e., $z(t)=\cos t, z^{\dagger}(t)=\sin t$. 
Then, the evolution of the partial observation $z(t)=\cos t$ is time-variant while that of the entire time series $\bs x(t)$ is time-invariant (as $\bs x(t+\interval)=B\bs x(t)$ for some rotation matrix $B=B(\interval)$). 
\end{ex}

\section[ARS Model]{Autoregressive with Slack Time Series (ARS) Model}
\label{sec:ARS}

Section~\ref{subsec:ARS} describes the proposed ARS model, and 
Section~\ref{subsec:estimation} describes the estimation procedure of the ARS model.

\subsection{ARS model}
\label{subsec:ARS}

To address the problems described in Section~\ref{subsec:problem_setting}, this study proposes \emph{autoregressive with slack time series~(ARS)} model, whose definition is based on a simple idea. In this model, we simply impute the missing variable $\{z^{\dagger}(jh)\}_{j=1}^{n}$ in AR model by a slack time series $z^{\dagger}_j$ to be estimated from the (partial) observations. Using the completed variables $x^{\ddagger}(jh)=(z(jh),z^{\dagger}_j)$, ARS model is then defined by
\begin{align}
    \hat{z}((j+1)\interval)
    =
    E_{r,s}
    \hat{B}\bs x^{\ddagger}(j\interval), \quad 
    \bs x^{\ddagger}(j\interval)
    =
    \left(
    \begin{array}{c}
        \bs z(j\interval) \\
        \hat{\bs z}^{\dagger}_j \\
    \end{array}
    \right),
    \label{eq:ARS}
\end{align}
where 
$E_{r,s}:=(I_r,O_{r,s})$ is a $r \times d$ matrix defined with the $r \times r$ identity matrix $I_r$ and the $r \times s$ zero matrix $O_{r,s}$. $\hat{B}$ is a matrix and $\hat{\bs z}^{\dagger}_j \in \mathbb{R}^{\tilde{s}}$ is a vector, where they are estimated by solving the following problem:
\begin{align}
    (\hat{B},\{\hat{\bs z}^{\dagger}_{j}\}_{j=1}^{n})
    &=
    \argmin_{(B,\{\bs z^{\dagger}_j\}_{j=1}^{n})}
    \ell(B,\{\bs z^{\dagger}_j\}_{j=1}^{n}), \text{ where} \nonumber \\
    \ell(B,\{\bs z^{\dagger}_j\}_{j=1}^{n})
    &:=
    \sum_{j=1}^{n-1}
    \|
        \bs x^{\ddagger}((j+1)\interval)
        -
        B \bs x^{\ddagger}(j\interval)
    \|_2^2.
    \label{eq:ARS_loss}
\end{align}
$\{\bs z^{\dagger}_j\}$ is especially called \emph{slack time series}, and the dimension $s \in \mathbb{N}$ of the vector $\bs z^{\dagger}_j$ is a user-specified parameter. 
The above optimization problem is solved with the aid of the simple linear regression analysis; see Section~\ref{subsec:estimation} for details.

While the idea behind the ARS model (i.e., the time series $\{\bs z^{\dagger}(j\interval)\}$ of missing variables is completed by the estimated slack time series $\{\hat{\bs z}^{\dagger}_j\}$) seems intuitive and simple enough, one natural question here is whether the completion really works. 
Regarding this question, for simplicity, 
this study assumes the time-invariance (and linearity) of the entire dynamical system to recover the entire dynamics from the partial observations. 
Then, we demonstrate the ARS model by a simple numerical experiment with missing variables shown in Figure~\ref{fig:illustration}. Therein, we first compute the dynamical time series $\bs x(t)=(z(t),z^{\dagger}(t))=(\cos t,\sin t)$ following the circular motion and assume that $r=s=1$ (i.e., $z(t)=\cos t$ is observed and $z^{\dagger}(t)=\sin t$ is missing) and $\interval=0.3,n=30$. 
The overall dynamics of the entire circular motion $\bs x(t)=(\cos t,\sin t)$ is time-invariant, while the dynamics of the partial observation $z(t)=\cos t$ is time-variant. 
Then, we forecast the future time series by leveraging AR(1) and ARS models (with $\tilde{s}=1$). 
Both AR(1) and ARS models are trained with only the observed time series $\{z(t)\}$ though ARS model additionally estimates the slack time series $\{z^{\dagger}(t)\}$. In this experiment of the circular motion, the slack time series is initialized by a standard normal random numbers, and is optimized by \verb|optim| function in \verb|R| language. 
The affirmative experimental result is also proved by our proposition; Section~\ref{subsec:discussion_theory} shows for the above simple case $d=2,r=s=1$ that the ARS model can recover the underlying true dynamics.

\bigskip

Historically speaking, it has been widely known that the time series of partially observed variables contain rich enough information to (partially) recover the entire dynamics. 
The well known Takens' theorem (also known as delay embedding theorem)~\cite{takens1981detecting} proves under some assumptions that the delay embedding $(z(t), z(t-\interval), z(t-2\interval),\ldots,z(t-(k-1)\interval)) \in \mathbb{R}^k$ computed only from the partial observation $z(t)$ has the same (topological) attractor structure to the entire dynamical time series $\{\bs x(t)\}$. Namely, the delay embedding at least roughly recovers the (topological) ``shape'' of the entire dynamical time series. 
While the classical Takens' theorem considers only the deterministic sequence, Takens' theorem can be further generalized to stochastic variants. See, e.g., \cite{baranski2020probabilistic}.

Although Takens' theorem only provides the rough estimate of the entire dynamical system through the delay embedding (computed only from the partial observations), we find that directly estimating the missing variables can improve the forecasting accuracy. 
Therefore, compared to previous approaches based on Takens' theorem, the simpler ARS model holds the potential to offer a more general and comprehensible model.

We last note that the minimizer of the ARS loss function $\ell(B,\{\bs z^{\dagger}_j\}_{j=1}^{n})$ is not unique. For instance, if we multiply $\bs z_j^{\dagger}$ by any positive real number $\alpha>0$, the matrix $B_{\alpha}$, whose corresponding rows are also multiplied by $1/\alpha$, yields the same loss function value 
$\ell(B,\{\bs z^{\dagger}_j\}_{j=1}^{n})=\ell(B_{\alpha},\{\alpha \bs z^{\dagger}_j\}_{j=1}^{n})$. There remain the freedom of the constant multiplication, though this multiplication does not affect the forecast, i.e., $\hat{\bs x}^{\ddagger}((j+1)\interval)=\hat{B}\bs x^{\ddagger}(j\interval)=\hat{B}_{\alpha}\bs x^{\ddagger}_{\alpha}(j\interval)$, where $\bs x^{\ddagger}_{\alpha}(j\interval)=(\bs z(j\interval)^{\top},\alpha \hat{\bs z}_j^{\dagger \top})^{\top}$. Also see Section~Section~\ref{subsec:discussion_theory} for our proposition, indicating the prediction uniqueness of the ARS model.

\subsection{Estimation of the ARS model}
\label{subsec:estimation}

While ARS model needs to estimate the AR model parameter $B \in \mathbb{R}^{d \times d}$ and the slack time series $\{\bs z^{\dagger}_j\}_{j=1}^{n}$ simultaneously, we can skip the estimation of the parameter $B \in \mathbb{R}^{d \times d}$. We describe the procedure in the following.

With given vectors $\{\hat{\bs z}^{\dagger}_j\}$ (whereby we obtain $\bs x^{\ddagger}(j\interval)=(\bs z(j\interval)^{\top},\hat{\bs z}^{\dagger\top}_j)^{\top}$), the AR model parameter $B \in \mathbb{R}^{d \times d}$ is estimated by a simple matrix formulae used in the linear regression analysis:
\[
    \argmin_{B \in \mathbb{R}^{d \times d}} \ell(B, \{\hat{\bs z}^{\dagger}_j\}_{j=1}^{n})
    =
    (\hat{D}^{\top} \hat{D})^{-1} \hat{D}^{\top}\hat{D}_+,
\]
where $\hat{D} \in \mathbb{R}^{(n-1) \times d}$ denotes the matrix concatenating the vectors $\bs x^{\ddagger}(j\interval) \in \mathbb{R}^d$ for $j=1,2,\ldots,n-1$, and 
$\hat{D}_+ \in \mathbb{R}^{(n-1) \times d}$ denotes the matrix of 1 step further of the time points, i.e., the matrix concatenating $\bs x^{\ddagger}((j+1)\interval) \in \mathbb{R}^d$. 
Note that the matrices $\hat{D},\hat{D}_+$ depend on the vectors $\{\hat{\bs z}^{\dagger}_j\}_{i=1}^{n}$. 
Then, the minimum loss function value is also obtained as
\begin{align}
    \min_{B \in \mathbb{R}^{d \times d}} \ell(B,\{\hat{\bs z}^{\dagger}_j\}_{j=1}^{n})
    =
    \text{tr}\{\hat{D}_+^{\top} (I-\hat{H}) \hat{D}_+\},
    \label{eq:partial_AR_minimum}
\end{align}
where $\hat{H}=\hat{D} (\hat{D}^{\top}\hat{D})^{-1}\hat{D}^{\top}$ is the hat matrix and $\text{tr}H=\sum_{j=1}^{d} h_{jj}$ denotes the trace of the matrix $H=(h_{jk})$, i.e., the sum of diagonal entries. 
Equation \eqref{eq:partial_AR_minimum} is obtained by $\hat{\bs x}^{\ddagger}=\hat{D}\hat{B}=\hat{D}(\hat{D}^{\top}\hat{D})^{-1}\hat{D}^{\top}\hat{D}_+=\hat{H}\hat{D}_+$ and $\hat{H}^2=\hat{H}$ (see, e.g., \cite{petersen2012matrix} for basic matrix formulae). 
As Equation~\eqref{eq:partial_AR_minimum} indicates that
\begin{align}
    \min_{(B,\{\bs z^{\dagger}_j\}_{j=1}^{n})} 
    \ell(B, \{\bs z^{\dagger}_j\}_{j=1}^{n})
    &=
    \min_{\{\bs z^{\dagger}_j\}_{j=1}^{n}}
    \min_{B \in \mathbb{R}^{d \times d}} 
    \ell(B, \{\bs z^{\dagger}_j\}_{j=1}^{n}) \nonumber \\
    &=
    \min_{\{\bs z^{\dagger}_j\}_{j=1}^{n}}
    \text{tr}\{\hat{D}_+^{\top} (I-\hat{H}) \hat{D}_+\},
     \label{eq:partial_AR_marginal_out}
\end{align}
in practice, we may solve the minimization problem \eqref{eq:partial_AR_marginal_out} by leveraging some general-purpose optimization functions. 
We use \verb|optim| function in \verb|R| language in our implementation.
Appendix~\ref{app:interaction} provides a possible extension of the ARS model to consider the interaction effects.

\section{Numerical Experiments}
\label{sec:experiments}

We examine AR and the proposed ARS models using synthetic datasets. 
Particularly, the experimental settings and results are shown in Section~\ref{subsec:settings} and \ref{subsec:results}, respectively. 
\verb|R| source codes to reproduce the experimental results are provided in \url{https://github.com/oknakfm/ARS}.

\subsection{Settings}
\label{subsec:settings}
\noindent
\textbf{Synthetic dataset generation:} 
we generate two different types of synthetic datasets, following (i) the circular motion and (ii) Lorenz dynamics. More specifically, (i) and (ii) are defined as follows.
\begin{enumerate}[{(i)}]
\item 
Circular motion: $\bs x(j)=(\cos (5+j/20), \sin (5+j/20))$. 

\item 
Lorenz dynamics: 
define the evolution function $g:\mathbb{R}^3 \to \mathbb{R}^3$ (of period of a constant length) for $\bs x=(x_1,x_2,x_3)$ as follows:
\begin{align*}
g(\bs x)&:=\Bigg(
    \Bigg(
    \begin{array}{cccccc}
            1 & 0 & 0 & 0 & 0 & 0 \\
            0 & 1 & 0 & 0 & 0 & 0 \\
            0 & 0 & 1 & 0 & 0 & 0 \\
    \end{array}
    \Bigg) \\
    &\hspace{1em}+ \dfrac{1}{200} \Bigg(
      \begin{array}{cccccc}
            -\alpha & \alpha & 0 & 0 & 0 & 0 \\
            \beta & -1 & 0 & 0 & -1 & 0 \\
            0 & 0 & -\gamma & 1 & 0 & 0 \\
      \end{array}
    \Bigg) \Bigg) (x_1,x_2,x_3,x_1x_2, x_1x_3, x_2x_3)^{\top},
\end{align*}
where $\alpha=10, \beta=28, \gamma=8/3$. 
The above $g$ is a first-order Taylor approximation of the original Lorenz dynamics~\cite{lorenz1963deterministic} shown in Equations~\eqref{eq:Lorenz1}--\eqref{eq:Lorenz3}. 
With the function $g^{(m)}$, which is the composition of the function $g$ of degree $m \in \mathbb{N}$, define $\bs x(0)=g^{(100)}(1/4,1/4,1/4)$ and $\bs x(j+1)=g(\bs x(j))$. 
\end{enumerate}

For the training set, we generate $10$ instances of the sequence of length $n=100$ for settings (i) and (ii). 
Independent normal errors with the standard deviations $\sigma=0,0.01$ are incorporated to the training sequences. 
For the test set (of forecasting), we compute the subsequent sequence of length $n=30,100$, for settings (i) and (ii), respectively.

\bigskip
\noindent 
\textbf{Missing mechanism:} 
for the setting (i), 
we regard the first $r=1$ entry as the observed variable: $z(j)=x_1(j)$, and the 
remaining $s=1$ entry as the missing variable: $z^{\dagger}(j)=x_2(j)$. 
For the setting (ii), we regard the first $r=2$ entry as the observed variable: $\bs z(j)=(x_1(j),x_2(j))$, and the remaining $s=1$ entry as the missing variable: $z^{\dagger}(j)=x_3(j)$.

\bigskip
\noindent
\textbf{Methods to be computed:}
using the training sets (of observed variables), we compute AR and ARS models. 
For ARS models, we employ $\tilde{s}=1$; 
we initialize the parameter by adding the standard Gaussian noise to the true missing parameter $z^{\dagger}(j)$. 
We employ \verb|optim| function with \verb|BFGS| option in \verb|R| package, to train the ARS models. 
While the slack time series here is initialized by adding random noise to the true missing parameters (for computational stability purposes), note that the series is initialized randomly by a standard normal distribution (without referencing the true parameters), in the forecasting of the circular motion shown in Figure~\ref{fig:illustration}.

\subsection{Results} 
\label{subsec:results}
Using the training instances, we train AR and ARS models, and forecast $\bs z(n+k)$, for $k=5,10,\ldots,25$. For the estimator $\hat{\bs z}(n+k)$ for each method, we compute the MSE $\hat{e}_k:=\|\bs z(n+k)-\hat{\bs z}(n+k)\|_2^2/r$ for $k=5,10,\ldots,25$. 
We further compute the relative error of ARS models to the baseline (AR model), and compute the mean and the standard deviation. 
The results for the settings (i) and (ii) are shown in Tables~\ref{table:Circular}--\ref{table:Lorenz}. 
We can observe that all the relative errors are less than $1$, i.e., the MSE of ARS models are smaller than that of the conventional AR models.

\begin{figure*}[p]
\centering
\begin{minipage}{0.48\textwidth}
\centering
\includegraphics[width=0.9\textwidth]{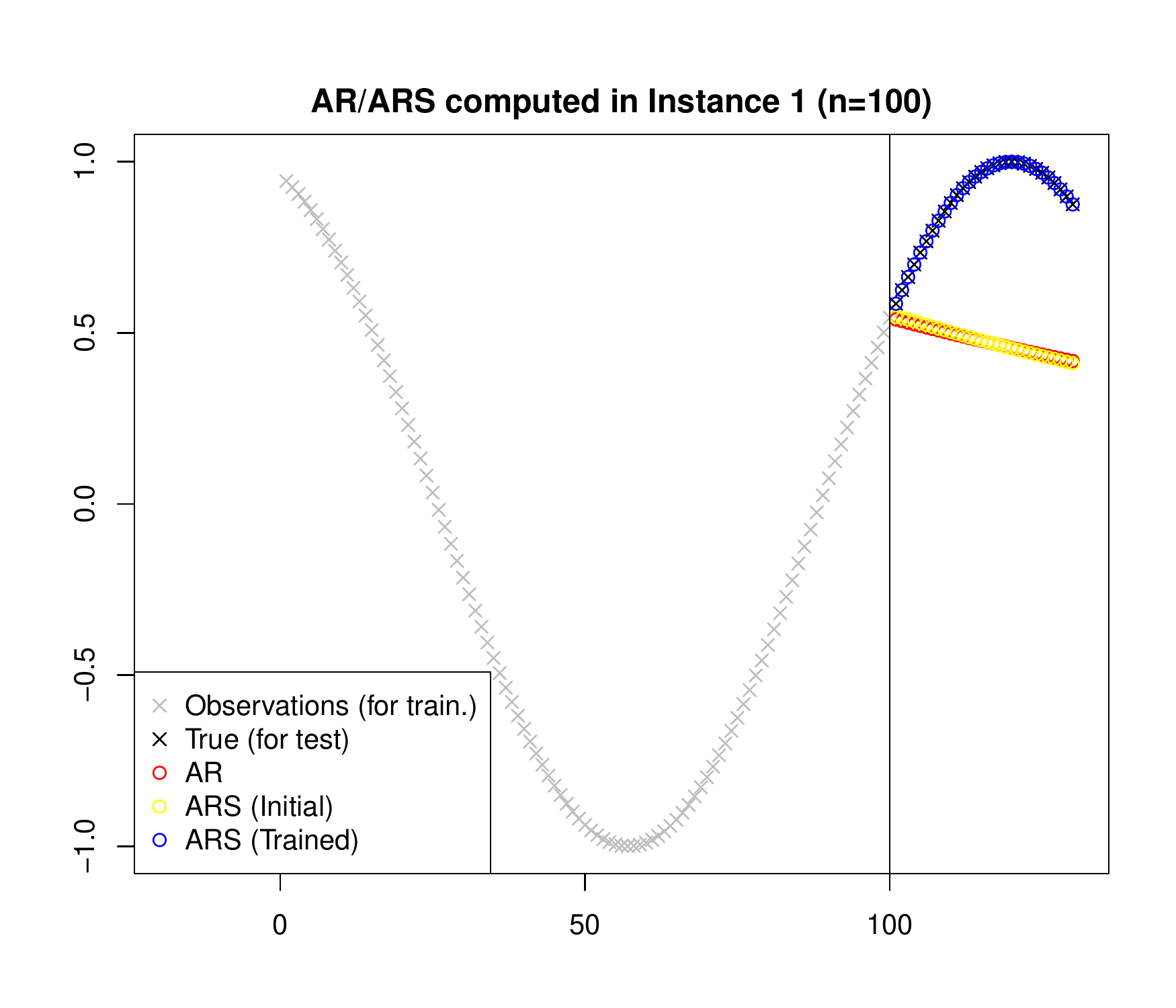} 
\caption{(i) Circular motion, $n=100, \sigma=0$}
\label{fig:Circular1}
\end{minipage}
\begin{minipage}{0.48\textwidth}
\centering
\includegraphics[width=0.9\textwidth]{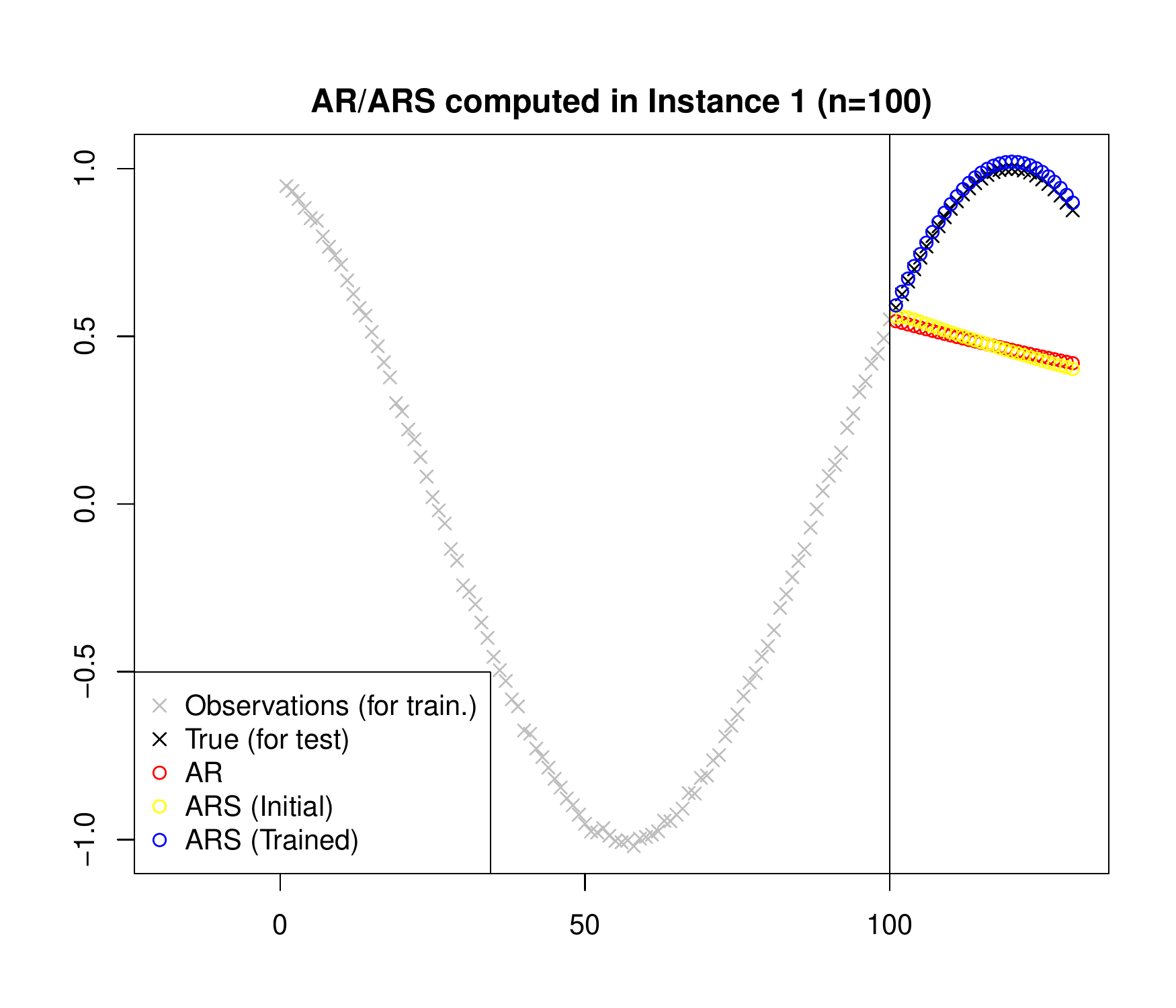} 
\caption{(i) Circular motion, $n=100, \sigma=0.01$}
\label{fig:Circular2}
\end{minipage}
\end{figure*}

\begin{figure*}[p]
\centering
\begin{minipage}{0.48\textwidth}
\centering
\includegraphics[width=0.9\textwidth]{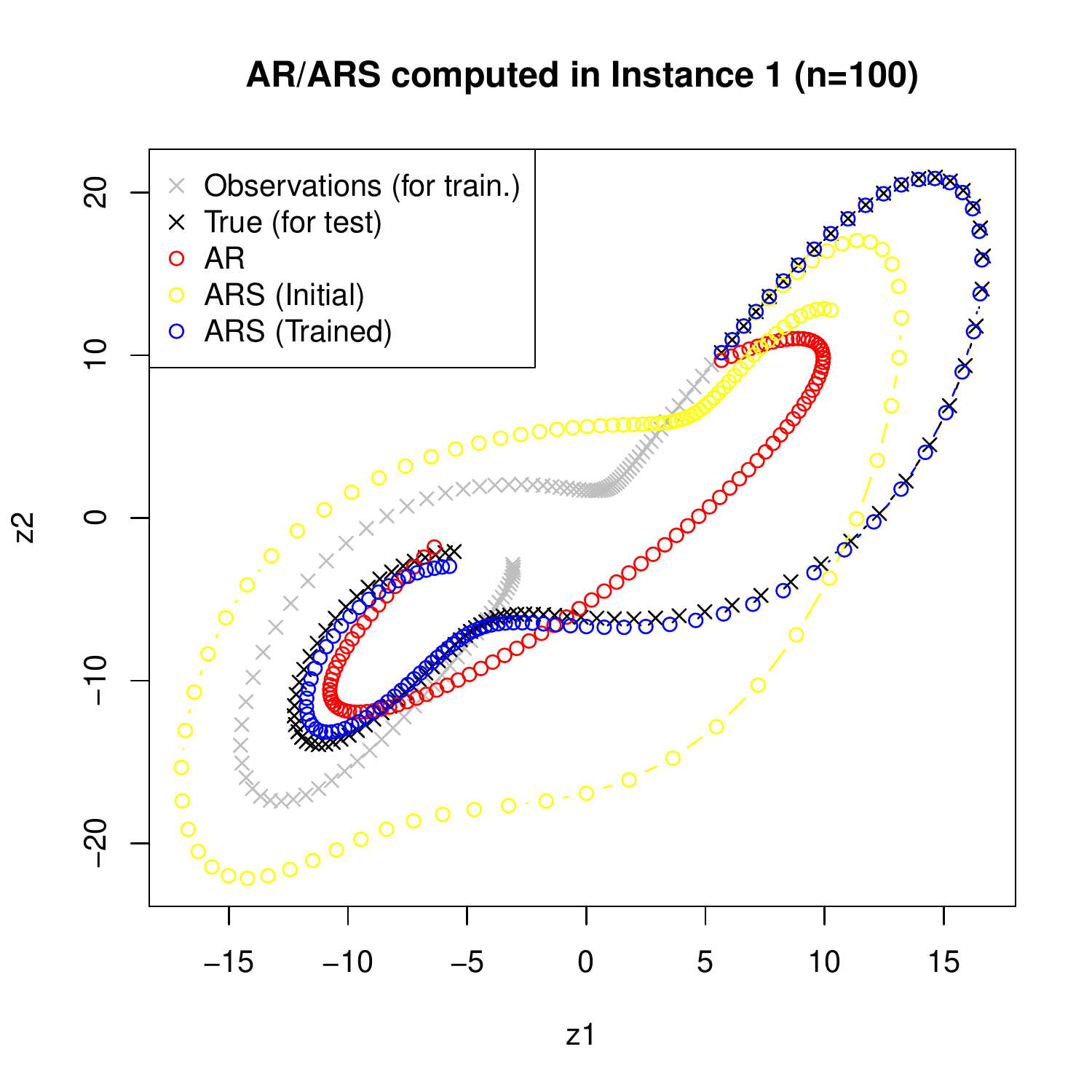} 
\caption{(ii) Lorenz, $n=100, \sigma=0$} 
\label{fig:Lorenz1}
\end{minipage}
\begin{minipage}{0.48\textwidth}
\centering
\includegraphics[width=0.9\textwidth]{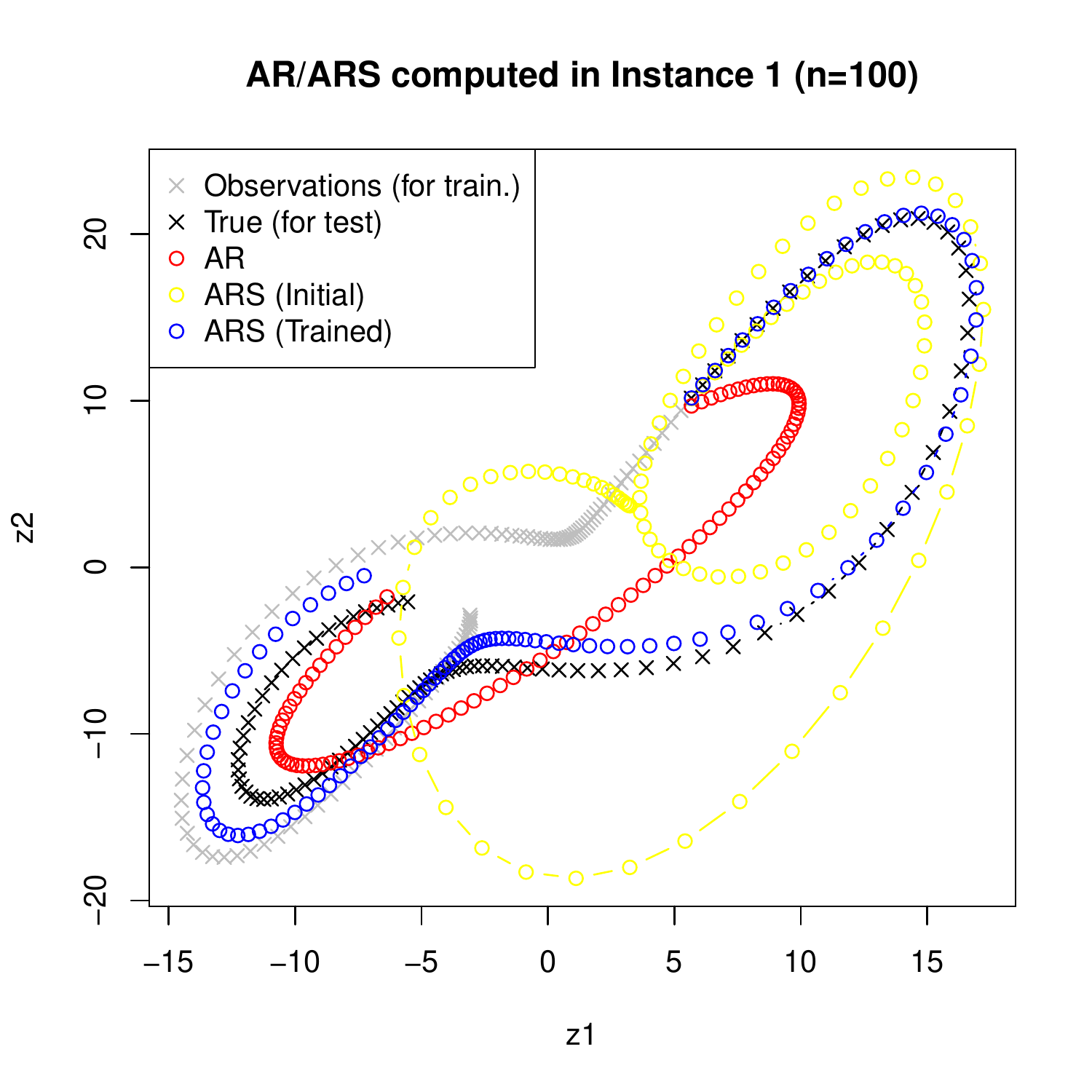}
\caption{(ii) Lorenz, $n=100, \sigma=0.01$}
\label{fig:Lorenz2}
\end{minipage}
\end{figure*}

For visualization, the training sets and the forecasts of AR/ARS models (using the training instance $1$) are also plotted in Figures~\ref{fig:Circular1}--\ref{fig:Lorenz2}. 
We can examine that ARS models forecast the true points better than the conventional AR model.

\begin{table*}[htbp]
\centering
\caption{Relative errors to AR model, with the setting (i) circular motion.}
\label{table:Circular}
\begin{tabular}{lcccccc}
\toprule
& $k=5$ & $k=10$ & $k=15$ & $k=20$ & $k=25$ & \\
\hline
$\sigma=0$ & $6.35 \pm5.37$ &
$7.26 \pm6.20$ &
$8.37 \pm7.28$ &
$9.82 \pm8.72$ &
$1.19 \pm1.09$ & $(\times 10^{-6})$
\\ 
$\sigma=0.01$ & 
$2.33\pm1.80$ &
$2.95\pm3.13$ &
$3.66\pm3.72$ &
$4.14\pm4.13$ &
$4.52\pm4.45$ & $(\times 10^{-1})$ \\
\bottomrule
\end{tabular}
\caption{Relative errors to AR model, with the setting (ii) Lorenz dynamics.}
\label{table:Lorenz}
\begin{tabular}{lcccccc}
\toprule
& $k=5$ & $k=10$ & $k=15$ & $k=20$ & $k=25$ & \\
\hline
$\sigma=0$ & 
$0.05\pm0.03$ &
$0.14\pm0.07$ &
$0.49\pm0.27$ &
$2.05\pm1.26$ &
$6.62\pm3.86$ & ($\times 10^{-2}$) \\
$\sigma=0.01$ & 
$0.13\pm0.06$ &
$0.15\pm0.10$ &
$0.28\pm0.20$ &
$0.77\pm0.58$ &
$1.77\pm1.47$ & ($\times 10^{-1}$)\\
\bottomrule
\end{tabular}
\end{table*}

\section{Discussions and Conclusion}
\label{sec:discussions_conclusions}

Section~\ref{subsec:discussion_theory}, \ref{subsec:discussion_statespace} and \ref{subsec:discussion_ARp} provide discussion on theory, and the relation to state-space model and higher-order AR model, respectively, and 
Section~\ref{subsec:conclusion} concludes this study.

\subsection{Discussion 1: ARS model can recover the underlying true dynamical time series}
\label{subsec:discussion_theory}

The ARS model involves a significant number of parameters: 
$\{z_j^{\dagger}\}_{j=1}^{n}$ contains $n \cdot s$ parameters, and 
$B$ contains $d^2$ parameters. Therefore, ARS model has a high degree of freedom (at least the ARS model is overparameterized, i.e., the number of parameters excees the sample size $n$), and it remains uncertain whether the ARS prediction is obtained correctly and uniquely. 

In the specific scenario of $d=2$, $r=s=1$, the answer to this question is affirmative. This section provides theoretical evidence that the ARS model can successfully recover the underlying true dynamical time series, regardless of the ostensibly large number of parameters.

To rigorously describe the theory, we employ the following notations: 
let $h>0$ be a fixed interval, and assume that $\bs x^{*\ddagger}_j=(z_j^{\top},z_j^{*\dagger\top})^{\top}$ with $z_j:=z(jh)$ follows an underlying true linear dynamics 
\begin{align}
    \bs x^{*\ddagger}_{j+1} = B^* \bs x^{*\ddagger}_{j}
    ,\quad (j=1,2,\ldots).
    \label{eq:underlying_true_dynamics}
\end{align}
Then, Proposition~\ref{PROP:PREDICTION} holds.

\begin{prop}
\label{PROP:PREDICTION}
Let $d=2,r=s=1$ and assume the identity \eqref{eq:underlying_true_dynamics} for the underlying true dynamics. 
With sufficiently large $n$, the future prediction via the ARS model: $\hat{z}_{n+k}=(1,0)\hat{B}^k\hat{\bs x}^{\ddagger}_{n}$ coincides with the underlying true $z_{n+k}$, for any $k \in \mathbb{N}$. 
\end{prop}

Proposition~\ref{PROP:PREDICTION} proves for the $2$-dimensional case (with $1$-dimensional observation) that the ARS model can recover the underlying true dynamical time series. See Appendix~\ref{app:proof_of_prop_prediction} for the proof. 
While we skip generalizing the proposition for simplicity, 
we can expect that the similar holds for $d>2$ by considering the numerical results shown in Section~\ref{sec:experiments}.

\subsection{Discussion 2: Relation to state-space model}
\label{subsec:discussion_statespace}

While the aforementioned experiments consider a small random error (where $\sigma=0,0.01$), we may employ a state-space model to deal with more influential random errors. 
A state-space model for linear systems is defined as 
\begin{align*}
    \bs z((j+1)\interval)
    &\sim 
    N(E_{r,s} \bs x((j+1)\interval), \Sigma_Z),\\
    \bs x((j+1)\interval)
    &\sim 
    N(B \bs x(j\interval), \Sigma_X),
\end{align*}
where $\{\bs x(j\interval)\}_{j=1}^{n} \subset \mathbb{R}^d$ represents a series of latent states, $E_{r,s} \in \mathbb{R}^{r \times d}$ represents the observation matrix, and 
$B \in \mathbb{R}^{d \times d}$ denotes the transition matrix. 
The proposed ARS model corresponds to the above state-space model defined with the observation matrix $E_{r,s}=(I_r,O_{r,s})$ where 
$I_r$ denotes the $r \times r$ identity matrix and $O_{r,s}$ denotes the zero matrix of size $r \times s$. 
While the state-space model is defined as a general framework, in practical situations, most of studies estimates only the hidden states $\bs x$ by preliminarily fixing the model parameter B. See, e.g., Bayesian filters including Kalman filters~\cite{kalman1960new,kalman1961new}. 
The difference to our ARS model is that the model parameter $B$ is usually fixed in most of the Bayesian filters, while ARS estimates both the unobserved variables and the model parameter $B$ simultaneously.

From another perspective, hidden markov model (HMM)~\cite{baum1966statistical} has much in common with the state-space model. However, in most cases, HMM employs a finite set for the state space $\setX$, and estimate the transition probability from a state $\bs x \in \setX$ to another state $\bs x' \in \setX$. 
Baum-Welch algorithm~\cite{BW1996} (also known as EM algorithm) assumes a latent structure on the hidden state $\bs x$, and taking marginal with respect to the unobserved $\bs x$. 
Viterbi algorithm~\cite{viterbi1967error} is closer approach to ours; it estimates the actual value of hidden state $\bs x$ by leveraging dynamic programming. 
While the state space $\setX$ in HMM can be extended to a set of countably many states~\cite{beal2001infinite}, HMM still focuses on modeling the transition of discrete states, while this paper considers a set of uncountably many states.

\subsection{Discussion 3: Relation to higher order AR model}
\label{subsec:discussion_ARp}

While this study so far considers AR(1) model $\hat{z}((j+1)h)=\hat{c} z(jh)$, we may employ higher order AR models. For instance, AR(2) model $\hat{z}((j+1)h)=\hat{c}_1 z(jh) + \hat{c}_2 z((j-1)h)$ is expected to have higher expressive power than the AR(1) model. This observation is true. 
Intuitively speaking, AR(1) model can approximate first-order differential equation by considering the relation $\diff z(t)/\diff t \big|_{t=jh} \approx \{z((j+1)h)-z(jh)\}/h$ with small $h>0$, while 
AR(2) model can approximate second-order one by considering the second-order central 
$\diff^2 z(t)/\diff t^2 \big|_{t=jh} \approx \{z((j+1)h)-2z(jh)+z((j-1)h)\}/h^2$. 
However, even increasing the order $p \in \mathbb{N}$, AR model is limited to approximate the dynamics described by a linear differential equation $\sum_{k=0}^{p}c_k\diff^k z(t)/\diff t^k=0$ (for some $c_0,c_1,\ldots,c_p \in \mathbb{N}$).
For instance, AR model cannot approximate the partial observation $x_1(t)$ of Lorenz system
\begin{align}
    \diff x_1(t)/\diff t 
    &=
    -\alpha x_1(t) + \alpha x_2(t), \label{eq:Lorenz1} \\
    \diff x_2(t)/\diff t 
    &=
    -x_1(t) x_3(t) + \beta x_1(t) - x_2(t), \label{eq:Lorenz2} \\
    \diff x_3(t)/\diff t 
    &=
    x_1(t) x_2(t) - \gamma x_3(t). \label{eq:Lorenz3} 
\end{align}
defined with parameters $\alpha,\beta,\gamma \ge 0$. This is because the partial observation $x_1(t)$ of the Lorenz system shown in Equations~\eqref{eq:Lorenz1}--\eqref{eq:Lorenz3} follows a non-linear differential equation. 
See Proposition~\ref{PROP:LORENZ} whose proof is shown in Appendix~\ref{app:proof_of_prop_Lorenz}. 

\begin{prop}
\label{PROP:LORENZ}
Partial observation $x_1(t)$ of Lorentz dynamics (shown in Equations~\eqref{eq:Lorenz1}--\eqref{eq:Lorenz3}) follows a non-linear ordinary differential equation. 
More specifically, $x_1(t)$ satisfies $\sum_{k=0}^{3} P_k(x_1(t)) \diff^k x_1(t)/\diff t^k=0$ for some functions $P_1,P_2,P_3,P_4$ and at least one of $P_1,P_2,P_3,P_4$ is a non-constant function. 
\end{prop}

Mathematically speaking, higher order AR model is a special case of the ARS model. 
Here, consider arbitrary time series $\{z((jh)\}_{j=1}^n$; this inclusion relation can be proved by substituting the delayed series $z((j-1)h)$ into the slack time series $\hat{z}^{\dagger}_j$ in ARS model. 
Then, the ARS model reduces to the AR(2) model $\hat{z}((j+1)h)=\hat{c}_1 z(jh)+\hat{c}_2 z((j-1)h)$. 
General AR($p$) model also can be implemented by substituting the delayed series $z((j-1)h),z((j-2)h),\ldots,z((j-p+1)h)$ to the ARS model in the same way. 
Furthemore, the ARS model (more rigorously speaking, extension of ARS model shown in Appendix~\ref{app:interaction}) is capable of representing the Lorenz equation while the higher-order AR model cannot describe Lorenz dynamics. 
We last note that the slack time-series can be incorporated to the higher-order AR models; we can easily define higher-order ARS models (though we skip the details in this paper for simplicity).

\subsection{Conclusion}
\label{subsec:conclusion}

This study considered the setting that some variables in dynamical time series were missing; we extended the autoregressive~(AR) model to propose AR with slack time series (ARS) model. 
The effectiveness of the ARS model was demonstrated by numerical experiments.

As this study provides the concept of the slack time series, there remain following limitations indicating potential future works. 
\begin{enumerate}
\item The optimization procedure is currently lacking in efficiency. 
We may consider more efficient and stable optimization algorithms; one possible approach is to employ some parametric models (or some restrictions) for the slack time series $\bs z^{\dagger}(t)$. 
\item Random errors included in more practical dynamical time series should be removed. We may combine the concept of slack time series to the conventional filters for state space models. 
\item The proposed ARS model still offers considerable scope for further theoretical exploration. 
It would be worthwhile to elucidate mathematical conditions that the proposed ARS approximates the underlying dynamics from the partial observations. 
Also the evaluation of the spatial and temporal complexity is preferred. 
\item ARS model has a high expressive capability (at least ARS model generalizes general order AR model as described in Section~\ref{subsec:discussion_ARp}, and it includes $n \cdot s$ parameters therein). 
Given this complexity, the potential for overfitting warrants careful consideration. To address this, some strategies such as regularization appears essential for practical applications. 
\item Lastly, the ARS model has been evaluated solely through experiments on synthetic datasets. We plan to estimate the proposed ARS to solve more practical problems. As an end-goal in mind, this study has been started to model the time evolution of the temperature/density profiles of fusion plasmas. 
This real-world application would be a demonstration for the potential utility of the proposed ARS model.
\end{enumerate}

\section*{Acknowledgment}
We would like to thank the editor, the AE, and four anonymous reviewers for constructive comments and suggestions. 
We also thank Kohei Hattori for helpful discussions. 
This cooperative research was launched in the project ``the statistical and mathematical modeling for plasma physics and complementary plasma data'', formally supported by strategic research projects grant (2022-SRP-13) from research organization of information and systems. We would like to thank Sadayoshi Murakami, Masayuki Yokoyama, and Naoki Kenmochi for helpful discussions in this project. We also thank the first director Kazuhei Kikuchi, who passed away young after the first meeting in 2022. We pray for his soul to rest in peace.

\appendices

\section{Extension: Interaction Effects}
\label{app:interaction}

As discussed in Section~\ref{subsec:dynamical_system} and \ref{subsec:autoregressive_model}, the AR model is derived from the short period approximation of the dynamical system (shown in Equation~\eqref{eq:short_period_approximation}) with the assumption $f(\bs x)=\diff \varphi(t,\bs x)/\diff t=A\bs x$ for some matrix $A \in \mathbb{R}^{d \times d}$. Unfortunately, however, this assumption $f(\bs x)=A\bs x$ is restrictive. For instance, a dynamical system following a Lorenz equation $f(\bs x)=(-\alpha x_1+\alpha x_2, -x_1x_3+\beta x_1-x_2, x_1x_2-\gamma x_3)$ for some constants $\alpha,\beta,\gamma \in \mathbb{R}$ (see \cite{strogatz2001nonlinear} for details) does not satisfy this assumption as interaction terms $x_1x_2$ and $x_1x_3$ are included therein. 

Therefore, to deal with the interaction effects in ARS model, we consider a function which generates interaction terms: 
\begin{align}
    \interaction(\bs x)
    &=(x_1,\ldots,x_d, x_1x_2, \ldots,x_1x_d, x_2x_3,\ldots,x_{d-1}x_{d}) \nonumber \\
    &:\mathbb{R}^d \to \mathbb{R}^{d(d+1)/2} 
    \label{eq:interaction}
\end{align}
for $\bs x=(x_1,x_2,\ldots,x_d) \in \mathbb{R}^d$, and define an extended ARS model:
\begin{align}
    \hat{\bs x}^{\natural}((j+1)\interval)
    =
    \hat{E} \interaction(\bs x^{\ddagger}(j\interval)),
\end{align}
where 
$\bs x^{\ddagger}(j\interval)$ is defined same as Equation \eqref{eq:ARS}. 
$\hat{E} \in \mathbb{R}^{d \times d(d+1)/2}$ is a matrix and $\hat{\bs z}^{\dagger}_j$ is a vector, where they are estimated by solving the following optimization problem:
\begin{align}
    (\hat{E},\{\hat{\bs z}^{\dagger}_{j}\}_{j=1}^{n})
    &=
    \argmin_{(E,\{\bs z^{\dagger}_j\}_{j=1}^{n})}
    \widetilde{\ell}(E,\{\bs z^{\dagger}_j\}_{j=1}^{n}), \nonumber \\
    \widetilde{\ell}(E,\{\bs z^{\dagger}_j\}_{j=1}^{n})
    &:=
    \sum_{j=1}^{n-1}
    \|
        \bs x^{\ddagger}((j+1)\interval)
        -
        E \interaction(\bs x^{\ddagger}(j\interval))
    \|_2^2.
    \label{eq:extended_ARS_loss}
\end{align}

While this study considers only the interaction of order $2$ for simplicity, the function~\eqref{eq:interaction} can be further generalized so as to include the interaction of order $3$ (i.e., $x_1x_2x_3$) and so on. 
Note that polynomial functions of sufficiently high degree can approximate any continuous functions (Weierstrass' theorem; see, e.g., \cite{jeffreys1988weierstrass}); the extended ARS model equipped with the interaction terms of sufficiently high degree is expected to express any time-invariant dynamical system (equipped with continuous and non-linear time derivative $f$).

\section{Proof of Proposition~\ref{PROP:PREDICTION}}
\label{app:proof_of_prop_prediction}

In this proof, $\bs e_1=(1,0)^{\top},\bs e_2=(0,1)^{\top}$ denote unit vectors. 
As the observations $\{z_j\}_{j=1}^{n}$ are assumed to follow the linear dynamics $\bs x_{j+1}^{\ddagger}=B \bs x_j^{\ddagger}$, we have the following identity: 
\begin{align}
    0
    &=
    \min_{B,\{z_j^{\dagger}\}}
    \sum_{j=1}^{n-1}\|\bs x_{j+1}^{\ddagger} - B\bs x_j^{\ddagger}\|_2^2 \nonumber \\
    &=
    \min_{B,z_1^{\dagger}}
    \sum_{j=1}^{n-1}
    \{z_{j+1} - \bs e_1^{\top}B^j\bs x_1^{\ddagger}\}^2,
    \label{eq:2dim_identiy}
\end{align}
where the last equality is obtained by substituting $z_{j+1}^{\dagger}=\bs e_2^{\top}B^j\bs x_1^{\ddagger}$. We prove the assertion by the following 2 steps.

\paragraph*{Step 1} 
In this first step, we show that the ARS model $\bs e_1^{\top}B^j \bs x_1^{\ddagger}$ is specified by identifiable few parameters. 
To this end, we employ the eigendecomposition $B=U^{-1}\Lambda U$, where $\Lambda$ is a diagonal matrix whose diagonal entries are $\lambda_1,\lambda_2 \in \mathbb{C}$, and $U$ is an eigen matrix. Then, we have
\begin{align*}
    \bs e_1^{\top} B^j \bs x_1^{\ddagger}
    &=
    \bs e_1^{\top} U^{-1} \Lambda^j U \bs x_1^{\ddagger} \\
    &=
    \frac{u_{22}(u_{11}z_1+u_{12}z_1^{\dagger})}{\Delta}
    \lambda_1^j -
    \frac{u_{21}(u_{21}z_1+u_{22}z_1^{\dagger})}{\Delta}
    \lambda_2^j \\
    &=
    \alpha \lambda_1^j - \beta \lambda_2^j,
\end{align*}
where $\Delta:=|U|=u_{11}u_{22}-u_{12}u_{21}$ and 
$\alpha:=u_{22}(u_{11}z_1+u_{12}z_1^{\dagger})/\Delta,
\beta:=u_{21}(u_{21}z_1+u_{22}z_1^{\dagger})/\Delta$. 
Considering that $u_{11},u_{12},u_{21},u_{22}$ are real numbers, 
and the eigenvalues $\lambda_1,\lambda_2$ of the matrix $U$ are compatible with $\{(u_{11}+u_{22}) \pm \sqrt{(u_{11}-u_{22})^2+4 u_{12}u_{21}}\}/2$, the following hold:
\begin{enumerate}
\item if $\lambda_1 \not \in \mathbb{R}$, we have $(\lambda_1,\lambda_2)=(\eta \exp(i \theta),\eta \exp(-i\theta))$ for some $\eta > 0$ and $\theta \in (0,2\pi) \setminus \{\pi\}$. 
\item if $\lambda_1 \in \mathbb{R}$, we have $(\lambda_1,\lambda_2)=(\eta_1,\eta_2)$ for some real numbers $\eta_1,\eta_2 \in \mathbb{R}$. 
\end{enumerate}
For the case 1), $\hat{z}_{j+1}=\bs e_1^{\top} B^j \bs x_1^{\ddagger}=2\alpha \eta^j \cos(j \theta)=:f^{(1)}_j(\alpha,\eta,\theta)$ as the imaginary part of $\alpha \lambda_1^j-\beta \lambda_2^j$ should be $0$. 
For the remaining case 2), $\hat{z}_{j+1}=\bs e_1^{\top} B^j \bs x_1^{\ddagger}=\alpha \eta_1^j -\beta \eta_2^j=:f^{(2)}_j(\alpha,\eta,\beta)$. 
Considering both cases 1) and 2), the essential parameters in the ARS prediction $\hat{z}_{j+1}$ are identifiable, i.e., 
(i) $f^{(1)}_j(\alpha,\eta,\theta)=f^{(1)}_j(\alpha',\eta',\theta')$ holds for all $j$ if and only if $(\alpha,\eta,\theta)=(\alpha',\eta',\theta')$, 
(ii) $f^{(1)}_j(\alpha,\eta,\theta)=f^{(2)}_j(\alpha,\eta,\beta)$ does not hold for all $j$, 
(iii) $f^{(2)}_j(\alpha,\eta,\beta)=f^{(2)}_j(\alpha',\eta',\beta')$ holds for all $j$ if and only if $(\alpha,\eta,\beta)=(\alpha',\eta',\beta')$.

\paragraph*{Step 2}
True parameters $B^*,\{z_j^{*\dagger}\}$ satisfy the equality \eqref{eq:2dim_identiy}. 
As the above Step 1 indicates that the ARS model is specified by identifiable few parameters, all the solutions satisfying the equality \eqref{eq:2dim_identiy} corresponds to the same (identifiable) parameters. 
Therefore, the prediction $\hat{z}_{n+k}$ is also uniquely determined and is compatible with the underlying true dynamical time series $z_{n+k}$. 
Therefore, the assertion is proved.

\qed

\section{Proof of Proposition~\ref{PROP:LORENZ}}
\label{app:proof_of_prop_Lorenz}

As Equation~\eqref{eq:Lorenz1} indicates that 
\begin{align}
x_2(t)=x_1(t)+\frac{1}{\alpha} \frac{\diff x_1(t)}{\diff t},
\label{eq:inter0}
\end{align}
we have 
\begin{align}
    \frac{\diff x_2(t)}{\diff t}
    =
    \frac{\diff x_1(t)}{\diff t}
    +
    \frac{1}{\alpha} \frac{\diff^2 x_1(t)}{\diff t^2}.
    \label{eq:inter1}
\end{align}
Substituting the identities~\eqref{eq:inter0} and \eqref{eq:inter1} into both sides of the equation~\eqref{eq:Lorenz2} yields 
\begin{align*}
    \frac{\diff x_1(t)}{\diff t}
    +
    \frac{1}{\alpha} \frac{\diff^2 x_1(t)}{\diff t^2}
    &=
    -x_1(t) x_3(t) + \beta x_1(t) - 
    \left\{
        x_1(t)+\frac{1}{\alpha} \frac{\diff x_1(t)}{\diff t}
    \right\}.
\end{align*}
Arranging the obtained terms proves
\begin{align}
    x_3(t) 
    &= 
    -\frac{1}{x_1(t)}
    \bigg\{
    \frac{1}{\alpha}
    \frac{\diff^2 x_1(t)}{\diff t^2}
    +
    \frac{1+\alpha}{\alpha}
    \frac{\diff x_1(t)}{\diff t} +
    (1-\beta)x_1(t) 
    \bigg\}.
    \label{eq:inter2}
\end{align}
This identity proves 
\begin{align}
    \frac{\diff x_3(t)}{\diff t}
    &=
    \frac{1}{x_1(t)^2}
    \left\{
    \frac{1}{\alpha}
    \frac{\diff^2 x_1(t)}{\diff t^2}
    +
    \frac{1+\alpha}{\alpha}
    \frac{\diff x_1(t)}{\diff t}
    +
    (1-\beta)x_1(t) 
    \right\} \nonumber \\
    &\hspace{3em} 
    -
    \frac{1}{x_1(t)}\bigg\{
    \frac{1}{\alpha}
    \frac{\diff^3 x_1(t)}{\diff t^3}
    +
    \frac{1+\alpha}{\alpha}
    \frac{\diff^2 x_1(t)}{\diff t^2} +
    (1-\beta)\frac{\diff x_1(t)}{\diff t}     
    \bigg\}. 
    \label{eq:inter3}
\end{align}
As the Lorenz equation~\eqref{eq:Lorenz3} indicates with Equations~\eqref{eq:inter1} and \eqref{eq:inter2} that 
\begin{align}
    \frac{\diff x_3(t)}{\diff t}
    &=
    x_1(t)x_2(t) - \gamma x_3(t) \nonumber \\
    &=
    x_1(t) 
    \left\{ 
    \frac{\diff x_1(t)}{\diff t}
    +
    \frac{1}{\alpha} \frac{\diff^2 x_1(t)}{\diff t^2}
    \right\}
    +
    \gamma 
    \frac{1}{x_1(t)}
    \bigg\{
    \frac{1}{\alpha}
    \frac{\diff^2 x_1(t)}{\diff t^2} +
    \frac{1+\alpha}{\alpha}
    \frac{\diff x_1(t)}{\diff t}
    +
    (1-\beta)x_1(t) 
    \bigg\}.
    \label{eq:inter4}
\end{align}

Comparing equations \eqref{eq:inter3} and \eqref{eq:inter4}, where both are multiplied by $x_1(t)^2$, yields
\begin{align*}
    &\left\{
    \frac{1}{\alpha}
    \frac{\diff^2 x_1(t)}{\diff t^2}
    +
    \frac{1+\alpha}{\alpha}
    \frac{\diff x_1(t)}{\diff t}
    +
    (1-\beta)x_1(t) 
    \right\} -
    x_1(t)
    \left\{
    \frac{1}{\alpha}
    \frac{\diff^3 x_1(t)}{\diff t^3}
    +
    \frac{1+\alpha}{\alpha}
    \frac{\diff^2 x_1(t)}{\diff t^2}
    +
    (1-\beta)\frac{\diff x_1(t)}{\diff t}     
    \right\} \\
    &=
    x_1(t)^3
    \left\{ 
    \frac{\diff x_1(t)}{\diff t}
    +
    \frac{1}{\alpha} \frac{\diff ^2 x_1(t)}{\diff t^2}
    \right\} +
    \gamma 
    x_1(t)
    \left\{
    \frac{1}{\alpha}
    \frac{\diff^2 x_1(t)}{\diff t^2}
    +
    \frac{1+\alpha}{\alpha}
    \frac{\diff x_1(t)}{\diff t}
    +
    (1-\beta)x_1(t) 
    \right\}.
\end{align*}
Therefore, rearranging the terms proves the assertion 
$\sum_{k=0}^{3} P_k(x_1(t)) \frac{\diff^k x_1(t)}{\diff t^k}=0$, where 
\begin{align*}
    P_0(x)
    &=
    (1-\beta)
    -
    (1-\beta) \gamma x, \\
    P_1(x)
    &=
    \frac{1+\alpha}{\alpha}
    -
    \left(
        \frac{1+\alpha}{\alpha} \gamma - (1-\beta) 
    \right) x
    -
    x^3, \\
    P_2(x)
    &=
    \frac{1}{\alpha}
    -
    \frac{1+\alpha+\gamma}{\alpha}x
    -
    \frac{1}{\alpha}x^3, \\
    P_3(x)
    &=
    -\frac{1}{\alpha}x.
\end{align*}

\qed

\end{document}